# Unconventional interfacial superconductivity in epitaxial Bi/Ni heterostructures

Xin-Xin Gong[1], Hexin Zhou[1], Peng-Chao Xu[1], Di Yue[1], Kai Zhu[1], Xiaofeng Jin[1*], He Tian[2], Gejian Zhao[3] & Ting-Yong Chen[3]

[1]State Key Laboratory of Surface Physics and Department of Physics, Fudan University, Shanghai 200433, China; Collaborative Innovation Center of Advanced Microstructures, Fudan University, Shanghai, 210433, China. [2]State Key Laboratory of Silicon Materials and School of Materials Science and Engineering, Zhejiang University, Hangzhou, Zhejiang 310027, China. [3]Department of Physics, Arizona State University, Tempe, AZ 85287, USA

*Corresponding author. Email address: xfjin@fudan.edu.cn

**Superconductivity (SC) is one of the most intriguing physical phenomena in nature. Nucleation of SC has long been considered highly unfavorable if not impossible near ferromagnetism (FM)[1-5], in low dimensionality[6,7] and, above all, out of non-superconductor. Here we report observation of SC with $T_C$ near 4 K in Ni/Bi bilayers that defies all known paradigms of superconductivity, where neither ferromagnetic Ni film nor rhombohedra Bi film is superconducting in isolation[8]. This highly unusual SC is independent of the growth order (Ni/Bi or Bi/Ni), but highly sensitive to the constituent layer thicknesses. Most importantly, the SC, distinctively non-$s$ pairing, is triggered from, but does not occur at, the Bi/Ni interface. Using point contact Andreev reflection (AR)[9], we show evidences that the unique SC, naturally compatible with magnetism, is triplet $p$-wave pairing. This new**

**revelation may lead to unconventional avenues to explore novel SC for applications in superconducting spintronics[10,11].**

Understanding how and why SC occurs in a given material has been very challenging for physicists for more than a hundred years, notwithstanding the major milestones, such as the London theory, the Landau-Ginzburg theory, and the BCS theory[12]. The extreme challenge to predict the occurrence of SC is symbolized by the long string of unanticipated but breathtaking advances – with the unexpected discoveries of cuprates and Fe-pnictides being the dramatic modern examples[13,14]. Because of their incompatibility, the nucleation of SC near a ferromagnet is difficult and has never been realized except cases where another superconductor provides proximity boosted cooper pairs[5,10,11]. This perceived necessity to start with another superconductor is engrained by the extensive study of the proximity effect in superconductor/ferromagnet (S/F) heterostructures, where all the structures involve a superconductor with either stable or metastable structure[15,16]. Compounding the difficulty, it is also generally recognized that SC with substantial $T_C$ is unfavorable in low dimensionality because of strong quantum fluctuation[6,7]. This paper reports a serendipitous finding of SC that emerges under the most implausible circumstances – near a ferromagnet, in low dimension, and in an otherwise non-superconducting material.

Bi in its thermodynamically stable rhombohedra structure is non-superconducting[8], which can be seen from the temperature dependence of the electrical resistance ($RT$ curve) of an epitaxial 20 nm Bi(110) film on Cu(001)/MgO(001), as shown by the yellow curve in Fig. 1a. On the other hand, Ni is a ferromagnetic element and has shown no traces of SC down to any measurable temperature. However, after a 3 nm Ni is deposited on Bi, the Ni(3 nm)/Bi(20 nm) becomes superconducting with $T_C$ about 4 K, as shown by the blue curve together with the inset in Fig. 1a. It should be noted that with the appearance of superconductor in Ni(3 nm)/Bi(20 nm), the Ni layer still remains ferromagnetic even above 300 K as shown by the magneto-optical Kerr effect (MOKE)[17,18] result in Fig. 1b. The incompatibility of FM and SC notwithstanding, this is the most dramatic nucleation of SC, in which a non-superconducting material becomes superconductor when it is covered by a strong ferromagnet.

Ni in its thermodynamically stable face-centered-cubic (fcc) structure is ferromagnetic but non-superconducting. Now starting from MgO(001) substrate, we epitaxially grow 3 nm Ni(001) film at 300 K[19] then cool down to 110 K and epitaxially grow 20 nm rhombohedra Bi(110) film on the top, and the good quality of the heterostructure with sharp interfaces is revealed by the reflection high-energy electron diffraction (RHEED) patterns and the cross section scanning transmission electron microscopy (STEM) in Fig. 1c. Strikingly, this Bi(20 nm)/Ni(3 nm) also becomes superconducting as shown by Fig. 1d; meanwhile the Ni layer underneath is clearly ferromagnetic as shown by Fig. 1e. It should be noted that our results are distinctively different from the previous report of S/F proximity effect in Bi/Ni bilayer where SC is originated from the novel fcc structure of Bi[15,16]. Since neither ferromagnetic Ni nor rhombohedra Bi is superconducting in isolation, and furthermore the SC is found to be independent of the deposition order of Ni and Bi, it is natural to conceive that the SC is induced at the interface. Indeed, the direct experimental evidence comes from the *RT* curve of Bi(6 nm)/Ni(0.18 nm)/Cu(2 nm) as shown in Fig. 1f, where 0.18 nm corresponds to one monolayer of

Ni. While SC has not fully developed, the onset of SC (blue dots) is unmistaken. In contrast, Bi(20 nm)/Cu(2 nm) is non-superconducting (yellow dots).

Even more striking is the thickness dependence of the SC in the Bi/Ni bilayers. We first fix the Bi film thickness at 15 nm and vary the Ni layer thickness to explore the influence of FM on the SC. For a small increasing thickness of Ni, the SC is substantially suppressed. As shown in Fig. 2a, for Ni layers between 2 – 4 nm thick, the SC is systematically suppressed and completely destroyed with 4 nm Ni. This result can be understood as a consequence of gradual penetration of the exchange interaction (effectively a strong magnetic field) from ferromagnetic Ni into the Bi film[5]. Because of the finite size effect on the Curie temperature (equivalently the exchange field) of an ultrathin ferromagnetic film, the exchange field at the interface of Bi/Ni should be significantly stronger when a Ni film changes from 2 nm to 4 nm thick[17,20,21]. Therefore when Ni films are very thin with sufficiently weak exchange interaction penetration, the Cooper pairs might survive near the Bi/Ni interface. Whereas, for thick Ni films, corresponding to strong exchange interaction penetration, the Cooper pairs are completely suppressed by the exchange field from the Ni layer, as shown schematically in Fig. 2d. These results clearly suggest the delicate dual role of the Ni layer; it creates the interface, which may trigger SC via yet to be identified microscopic mechanism (*e.g.*, electron-phonon or ferromagnetic fluctuation, *etc*), but it is also the detriment to eventually destroy the Cooper pairs.

Since the Bi layer is nonmagnetic and non-superconducting, there should be no effect on the SC from the Bi thickness. However, as shown next, the thickness of the Bi layer plays a crucial role for the nucleation of the SC. In Fig. 2b, we measured a series of Bi/Ni(4 nm) samples with a fixed Ni(4 nm) layer for which the SC has already been completely suppressed as shown in Fig. 2a. As we increase the Bi layer from 15 nm, remarkably the SC reappears again with the same $T_C$ of 4 K. Since the SC in the Bi(15 nm)/Ni(4 nm) is completely suppressed by the exchange field from the Ni layer, further increasing the Bi layer thickness cannot modify the interface or affect the Ni layer, the recurrence of the SC implies that the SC does not occur at the interface even though it may be triggered by the interface, which is again diametrically different from the S/F proximity effects. In fact the reappearance of the SC is a salient feature for the Ni/Bi superconducting systems. As seen in Fig. 2c, when $T_C$ is plotted as a function of the Bi film thickness for several Ni film thicknesses, in each case there is always a similar sharp increase (within 5 nm) to saturation. For example, the SC for Ni(6 nm) reappears at about 30 nm of Bi, very far from the interface, and its $T_C$ reaches about 4 K at 35 nm of Bi. This result also excludes unambiguously the possibility that the SC in Bi/Ni comes from Bi-Ni alloy if any at the interface.

Based on the foregoing results, we construct a series of schematics in Fig. 2d, e and f, to illustrate the occurrence and evolution of SC in the Bi/Ni system. This unusual SC behavior in Bi/Ni must be related to the unique electronic property of the Bi thin film. It has been realized that the interior of Bi thin films up to 90 nm is semiconducting (also confirmed here in this work by the yellow curve in Fig. 1a) then becomes semimetal for thicker films; meanwhile all the surfaces of Bi films including the top, bottom and side surfaces are always metallic[22,23], irrespective of the orientations of the exposed surfaces[23,24]. It is also known that, at 4 K, the thermal electronic excitation in the film interior appears already as the Bi film is thicker than 20 nm. Because of the well-known unusually long Fermi wavelength (~ 30 nm) and extremely long mean free path of electrons (up to mm) in Bi, as well as the extremely long decay of the inter-surface interaction in Bi films[25-28], the Cooper pairs could in principle be triggered via the interface in various channels but away from the interface, *e.g.*, two electrons paired in the film interior or on the opposite surface to the interface, as shown schematically in Fig. 2f. In fact, this is distinctively different from any other superconducting systems where Cooper pairs always live at the place of the cause of interaction. These aspects suggest that the highly unusual SC in epitaxial Bi/Ni, differing from those of all known superconductors, may be topological in nature.

Since a point contact only measures the resistance at the vicinity of the contact, the contact resistance between a superconductor and a normal metal depends strongly on the contact size and can vary from a few $\Omega$ to over 1000 $\Omega$. However, the contact resistance between two superconductors does not depend strongly on the contact size and this can be used to verify the SC at the surface of the Bi/Ni bilayers. We choose Indium, which is a conventional superconductor with $T_C$ = 3.41 K, as a soft tip to contact on the bare sample Bi (20 nm)/Ni (2 nm). A schematic setup of the point contact is illustrated in the inset of Fig. 3a. We have made over 10 contacts, all the contacts shows a similar resistance very close to zero, showing that the surface of Bi/Ni bilayer is indeed superconducting. The temperature dependence of a representative contact resistance is shown in Fig. 3a. The $T_C$ of both superconductors are obvious and consistent with Fig. 1. This result shows that the Bi surface is indeed superconducting.

To further reveal the nature of the SC in the Bi/Ni system, we studied the samples using point contact AR[9]. For a conventional superconductor such as Nb, the Andreev spectrum shows the well-known double-peak structure in the differential conductance (*dI/dV*), as shown by the inset of Fig. 3b for a 100 nm Nb film measured with a gold tip at 4.2 K. The peak position indicates the gap value of Nb and the change of *dI/dV* (the ratio of *dI/dV* at zero bias to that at large bias) must be less than 2, the Andreev limit[29]. We have measured the Bi/Ni bilayer samples with and without a 2 nm gold protection layer using materials of zero spin polarization such as gold and highly spin-polarized materials such as $La_{0.67}Sr_{0.33}MnO_3$ (LSMO). We have measured over 100 contacts on these samples with contact resistance from a few Ω to over 1000 Ω for both LSMO and gold tips. Most remarkably, there is only a single peak at zero bias. None of the Andreev spectra exhibit the double-peak feature as that of all conventional superconductors (*e.g.*, Nb) and Fe pnictides. One representative Andreev spectra is shown in Fig. 3b for a gold tip in contact on the sample Bi(20 nm)/Ni(2 nm)/MgO(100). The change of the *dI/dV* peak can be much larger than the Andreev limit 2 and a highly spin-polarized current from the LSMO tip cannot suppress the peak, as shown in Fig. 3c. These results rule out conventional *s*-wave pairing. A possibility could be SC naturally compatible with magnetism, such as triplet *p*-wave pairing. Finally, it should be mentioned that the in-plane upper critical field $H_{c2}$ is beyond the Pauli limit (see Extended Data and Supplementary Information), which is consistent with foregoing AR results.

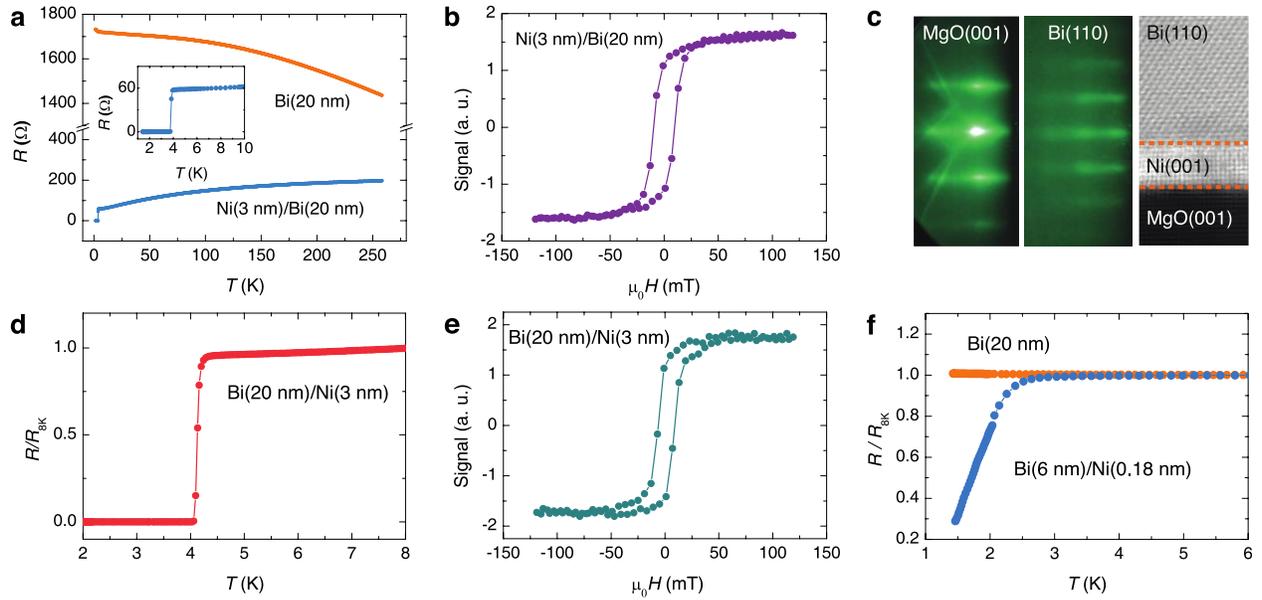

**Figure 1 | Observation of SC and FM in both epitaxial Ni/Bi and Bi/Ni heterostructures. a,** The *RT* curves of Bi and Ni/Bi on Cu(2nm)/MgO(001). Inset is the low temperature part of blue curve. **b,** In situ MOKE for Ni/Bi measured at 300 K. **c,** RHEED and cross section STEM of Bi(110)/Ni(001)/MgO(001). **d,** The *RT* curve of Bi/Ni/MgO. **e,** In situ MOKE for Bi/Ni measured at 300 K. **f,** *RT* curves of Bi/Cu/MgO sample and Bi/Ni/Cu/MgO sample.

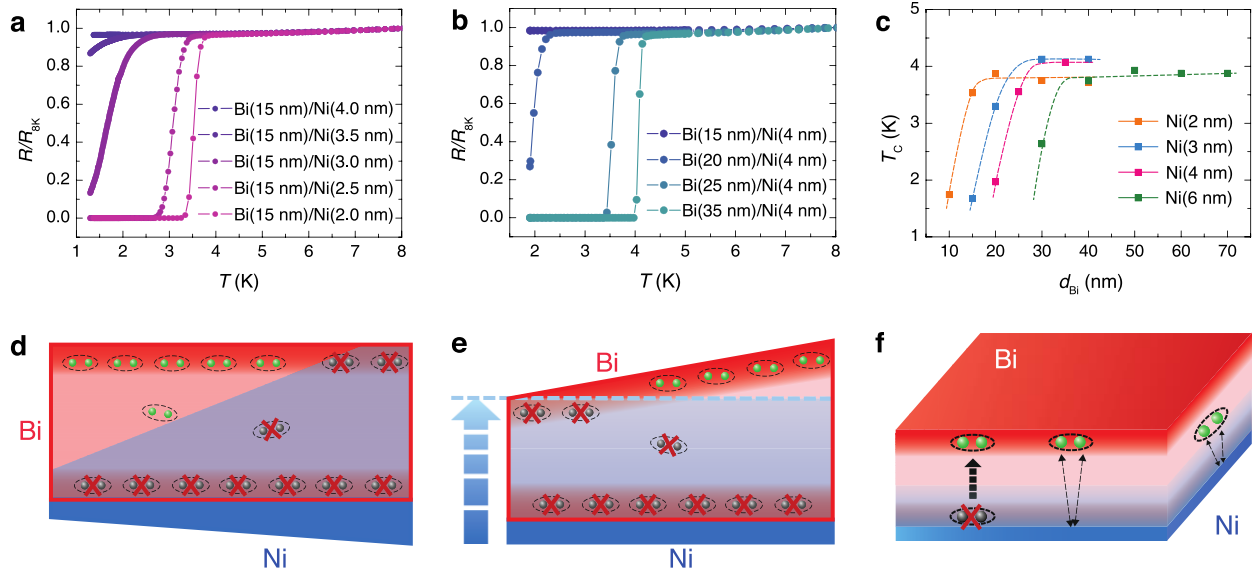

**Figure 2 | Both Ni and Bi thickness dependence of SC. a,** SC with variable Ni thicknesses. **b,** SC with variable Bi thicknesses. **c,** $T_C$ versus Bi thickness for different Ni thickness. **d–f,** Schematic illustration of the occurrence and evolution of SC in the Bi/Ni system.

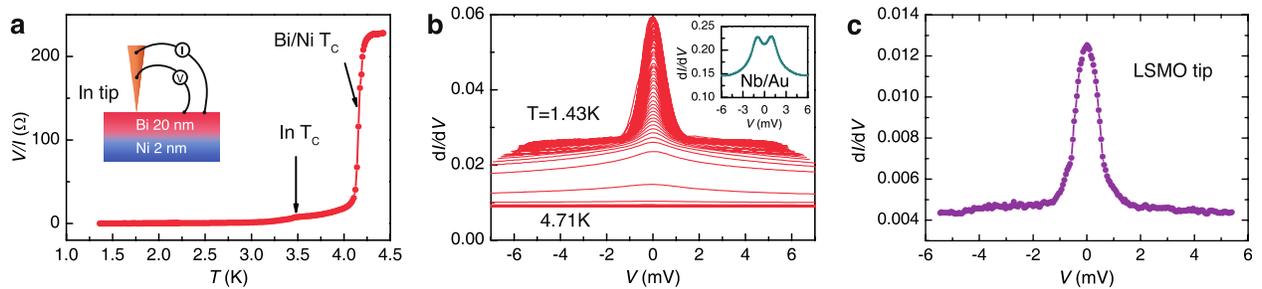

**Figure 3 | Point contact measurements on Bi(20 nm)/Ni(2 nm). a,** Resistance of a point contact using an Indium tip. **b,** Andreev spectra of a point contact using a gold tip from 1.43 K to 4.71 K with a temperature step about 0.25 K. Inset: Andreev spectrum of a gold tip in contact with a Nb film of 100 nm. **c,** Andreev spectra of a point contact using a LSMO tip in contact with the Bi(20 nm)/Ni(2 nm).

**Supplementary Information** is available in the online version of the paper.

**Acknowledgements** We thank Drs. Chia-ling Chien, Donglai Feng, Dunghai Lee, Shi-Yan Li, Li Lu, Tao Xiang, Yayu Wang, Fuchun Zhang, and Guang-ming Zhang for their valuable comments and suggestions. This work was supported by MOST (Grant No. 2015CB921400, 2011CB921802) and NSFC (Grants No. 11374057, No. 11434003 and No. 11421404) in China

and the research of Andreev reflection was supported as part of the SHINES, an Energy Frontier Research Center funded by the U.S. Department of Energy, Office of Science under Award SC0012670.

**Author Contributions** X.-X.G., H.Z., P.-C.X., D.Y., K.Z. and X.J. prepared the samples and performed transport and MOKE measurements. H.T. performed the STEM measurements. G.Z. and T.-Y.C. performed the Andreev reflection experiments. All authors discussed the results and contributed to the writing of the manuscript.

**Author Information** Reprints and permissions information is available at www.nature.com/reprints. The authors declare no competing financial interests. Readers are welcome to comment on the online version of the paper. Correspondence and requests for materials should be addressed to X. J. (xfjin@fudan.edu.cn).

**Methods**

All the samples were epitaxially grown in ultrahigh vacuum (UHV) with base pressure $\sim 6\times10^{-8}$ Pa. The single crystal MgO(001) substrates were first chemically cleaned before putting into the UHV chamber, then annealed in UHV at 750 K for 70 minutes. For Ni/Bi/Cu/MgO samples, the deposition of Cu was carried out at 300 K, while both Bi and Ni layers at 110 K. For Bi/Ni/MgO samples, the Ni layer was deposited at 300K while the Bi layer at 110K. The deposition rate was determined by a quartz microbalance. To prevent oxidation, a 4 nm MgO capping layer would be deposited on every sample before it was taken out of the UHV chamber for ex situ transport measurement. The transport data was measured in an Oxford Cryo-free magnet system (9 T, 1.5 K).

Point contact Andreev reflection was carried out in a custom-built Janis system down to 1.5 K with a vector magnetic field of 4 T and a 9 T field in z-direction. The tip and the sample are enclosed into a vacuum jacket, then cooled down to the desired temperature. A point contact is established using a differential screw mechanism when the temperature is stable, then the experiments are carried out subsequently.

**Supplementary Information**

Quasi-2D characteristics of the Bi(15nm)/Ni(2nm)/MgO(001) sample

The field dependence of $RT$ curves in both perpendicular and in plane directions are provided in Extended Data Fig. 1a and b, respectively. From each of these curves, we can extract three characteristic temperatures: the onset temperature ($R=95\%R_N$), the mid-point temperature ($R=50\%R_N$) and the zero-resistance temperature ($R=5\%R_N$). By plotting them versus external fields, as shown in Extended Data Fig. 1c, we obtain the temperature dependence of the upper critical fields in both direction. As for the perpendicular case (open symbols), the temperature dependence is almost linear down to ~2.0 K. Using the Werthamer–Helfand–Hohenberg (WHH) formula, $H_{C2}^{\perp}(0) = 0.69 \mathrm{d}H_{C2}^{\perp}/\mathrm{d}T_C |_{T_C}$ [30], the upper critical field at zero temperature from the zero-resistance data (indicated by the empty blue triangle) is estimated to be close to 1.9 T. For the in plane case (solid symbols), it is seen that the experimental data can be well fitted by $H_{C2}^{//}(T) = H_{C2}^{//}(0)(1-T/T_C)^{\alpha}$, yet the fitting parameter $\alpha$ is roughly 2/3, which is quite different from 1/2 as found in other quasi-2D superconductors[31,32]. For the onset data (indicated by the solid purple triangle), as a comparison to our fitting curve (solid purple line) with $\alpha = 2/3$, we plot explicitly here the fitting curve (dashed purple line) with $\alpha = 1/2$; obviously they deviate significantly from each other. Based on our fitting result in Fig. S1c, it is also noticed that all the three intercepts of the upper critical field at zero temperature are well above the Pauli limit $B_{Pauli}=1.83T_C=7.1$(T) [33,34]. In addition, as expected from a quasi-2D superconductor[32,35], the ratio $H_{C2}^{//} / H_{C2}^{\perp}$ is indeed divergent when the temperature is on approaching $T_C$.

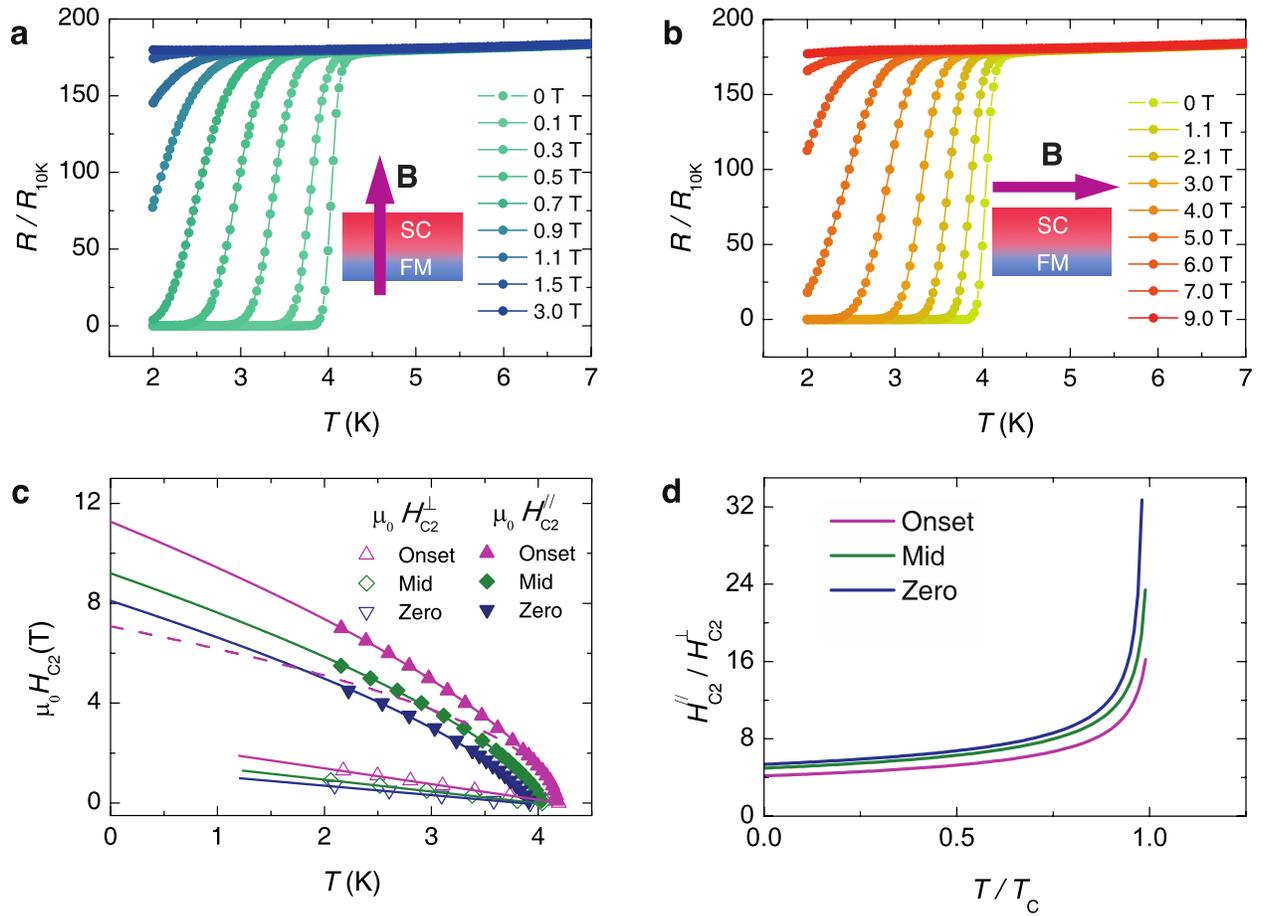

**Extended Data Figure 1 | Both out of plane and in plane $H_{C2}$ versus $T$ curves for Bi(15nm)/Ni(2nm)/MgO. a,** and **b,** The SC transition with perpendicular and in plane external magnetic field respectively. **c,** Temperature dependence of perpendicular (open symbols) and in plane (solid symbols) $H_{C2}$. The solid curves are the fittings using equation $H_{C2}^{//}(T) = H_{C2}^{//}(0)(1-T/T_C)^\alpha$, and the dashed yellow curve is the fitting from

$H_{C2}^{//}(T) = H_{C2}^{//}(0)(1-T/T_C)^{1/2}$. **d,** The divergent nature of the ratio $H_{C2}^{//}/H_{C2}^{\perp}$ on approaching $T_C$.